\documentclass[sigconf]{acmart}
\usepackage[utf8]{inputenc}

\usepackage{multicol}
\usepackage{graphicx}
\usepackage{epsfig}
\usepackage{multirow}
\usepackage{amssymb}
\usepackage{amsmath}
\usepackage{dsfont}
\usepackage{bbm}
\usepackage[ruled,linesnumbered]{algorithm2e}

\usepackage{mathtools}

\usepackage[left]{lineno}
\usepackage{blindtext}

\usepackage{ upgreek }
\usepackage{caption}
\usepackage{subfig}

\date{October 2018}

\usepackage{natbib}
\usepackage{graphicx}
\usepackage{booktabs} 

\setcopyright{rightsretained}

\makeatletter
\renewcommand\@formatdoi[1]{\ignorespaces}
\makeatother
\setcopyright{rightsretained}
\acmDOI{ }
\acmISBN{}
\setcopyright{rightsretained}
\acmConference[RecSysKTL'18]{Workshop on Intelligent Recommender Systems by Knowledge Transfer and Learning}{October 2018} {Vancouver, Canada}
\acmYear{2018}
\copyrightyear{2018}
\acmArticle{}
\acmPrice{}

\begin{document}
\title{Detecting Changes in User Preferences using Hidden Markov Models for Sequential Recommendation Tasks}
\renewcommand{\shorttitle}{Detecting Changes in User Preferences using HMMs for Sequential Recommendation}

\author{Farzad Eskandanian}
\orcid{1234-5678-9012}
\affiliation{%
  \institution{Center for Web Intelligence \\ DePaul University}
  \city{Chicago}
  \state{IL, USA}
}
\email{feskanda@depaul.edu}

\author{Bamshad Mobasher}
\affiliation{%
  \institution{Center for Web Intelligence \\ DePaul University}
  \city{Chicago}
  \state{IL, USA}
}
\email{mobasher@cs.depaul.edu}




\begin{abstract}
Recommender systems help users find relevant items of interest based on the past preferences of those users. In many domains, however, the tastes and preferences of users change over time due to a variety of factors and recommender systems should capture these dynamics in user preferences in order to remain tuned to the most current interests of users. In this work we present a recommendation framework based on Hidden Markov Models (HMM) which takes into account the dynamics of user preferences. We propose a HMM-based approach to change point detection in the sequence of user interactions which reflect significant changes in preference according to the sequential behavior of all the users in the data. The proposed framework leverages the identified change points to generate recommendations in two ways. In one approach change points are used to create a sequence-aware non-negative matrix factorization model to generate recommendations that are aligned with the current tastes of user. In the second approach the HMM is used directly to generate recommendations taking into account the identified change points. These models are evaluated in terms of accuracy of change point detection and also the effectiveness of recommendations using a real music streaming dataset. 
\end{abstract}

\maketitle

\section{Introduction}

Recommender systems learn from user profiles in order to find items that are of interest to users. Traditional approaches to recommendation involve modeling long-term preferences of users and tailoring the recommendations to the user based on those users' overall preference profiles. This approach, however, does not take into account the fact that a user's preference change over time and items that may have been relevant or of interest in the past may no longer suit the needs of the user. Changes in a user's preferences may occur because of changes in user's situation, context, the task at hand, or even due to one-time external events.

This problem is particularly pronounced in domains where a user's interest in items may vary often in the course of interactions with the system. An example of such a domain is music streaming where a user's sequence of interactions with the system (such as selecting, liking or disliking a song) is recorded and some or all of this history of interactions is used to identify future items to present to the user. But, even in domains with less transient behavioral characteristics, user preferences may change over time because a user's tastes may evolve slowly and thus older items in the user profile may no longer reflect the user's current preferences \cite{mcauley2013amateurs}. Examples of such a situation may be user preferences for wine where a user's taste may evolve over time due to experience and the development of a more discerning palate. 

To address this situation, recommender systems must be able to model the dynamics of user preferences based on historical data by identifying change points in the user interaction sequence with the system beyond which a user's behavior might indicate a significant change in preferences, and finally to tailor the system's recommendations to the most relevant episodes within the user's overall profile associated with the identified change points.  

Recent research in recommender systems has tried to address different aspects of this problem. For example, context-aware recommender systems (CARS) \cite{adomaviciusCARS2011, haririMusicContext2012, karatzoglou2010multiverse, zheng2014cslim} try to take into account the current context of the user or the most appropriate context for an item when generating recommendations. Most common approaches to CARS, however, rely on explicit representation of contextual factors which are not always available and generally do not try to model the dynamics of user preferences. Furthermore, session-based recommender systems \cite{jannach2017recurrent, session_hidasi2016parallel} have been introduced with a focus on developing models that generate recommendations using only the observed behavior of a user during an ongoing session while ignoring parts of user's overall profile that are considered to be associated with previous sessions. While session-based recommenders address part of the aforementioned problem by trying to model sequential interactions with the system, they generally do not address the problem of explicitly modeling and detecting change points in user preference sequences. 




In this work we present a recommendation framework based on Hidden Markov Models (HMM) which integrates automatic change point detection within sequences of user interactions with  recommendation models that take into account the dynamics of user preferences. We specifically focus on a setting where user preferences are implicitly and sequentially captured during the course of a user's interaction with the system and where there is no explicit representation for contextual or other factors that may provide {\em a priori} indications of possible changes in user behavior. The goal of this framework is to effectively identify change points in user preferences and use the identified change points to generate sequential recommendation. We conjecture that appropriate identification of change points and tailoring recommendations to the most relevant segments in the history of user interactions will lead to more accurate and effective recommendation. 

While there are various sequential pattern mining and modeling approaches that can be used to predict next items in a sequence \cite{rendleFPMC2010}, HMMs are particularly well-suited for this problem. Given a sequences of user-item interactions, an HMM can be used to identify the most likely sequences of hidden states representing change points in user preferences. This change point detection mechanism can be used to identify specific segments of user-item interaction sequences to be used in as input for a traditional non-sequential recommendation model such as matrix factorization. This approach, in and of itself, should lead to more effective recommendation when compared to the same approach without change point detection. However, the advantage of using HMMs is that the same learned model can be used to infer probabilities associated with items in the observation sequences which in turn can be used to directly generate recommendations for next items without resorting to other non-sequential recommendation models.

\begin{figure}
    \centering
    \includegraphics[scale=0.20]{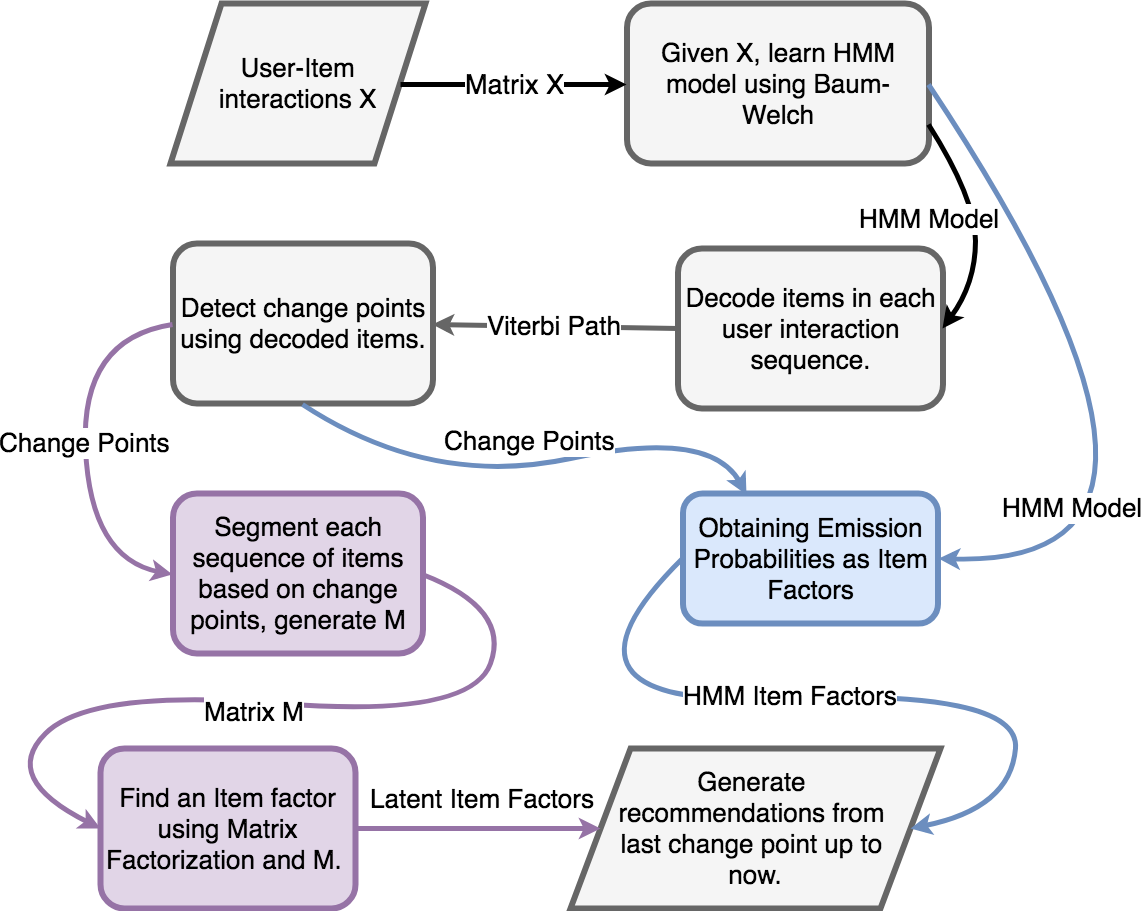}
    \caption{The process of our framework.}
    \label{fig:flow_diagram}
\end{figure}

The center piece of our proposed framework is a change detection mechanism using an HMM. We then integrate this mechanism into two separate recommendation models. In one approach we use the the change detection with standard non-negative matrix factorization. This allows us to determine whether the change detection can improve the effectiveness of recommendation when used in conjunction with standard non-sequential recommendation models. In the second approach, we develop a recommendation method using the HMM's sequential latent space model. In the latter case, the emission probabilities associated with items, together with the identified change points are used to directly generate recommendations. Figure~\ref{fig:flow_diagram} depicts the high-level process for the overall framework.

In this preliminary work, we empirically evaluate our approach using a real music streaming data set from Spotify, Inc. We compare the HMM-based change point detection with other standard baseline methods for identifying change points. We also compare the effectiveness of recommendations with and without change point detection using standard matrix factorization approach,  and we compare the proposed HMM-based sequential recommendation method to matrix-factorization with change point detection.

\section{Related Work}

There are three general areas of research in recommender systems that address the problem considering changes in user preferences, interests, or situations. One is Context-Aware Recommendation Systems (CARS) \cite{hariri2013query, adomaviciusCARS2011}. Most CARS approaches assume a predefined representational view to modeling context and the main goal is to leverage contextual information in generating recommendations. 

Some other works, has studied the case which the contextual information is not observable directly but we can infer them using sequential behavior of users in the system \cite{haririMusicContext2012}. However, these approaches are devised to infer the contextual information implicitly using topic modeling of content information such as tags associated with items. Also, in \cite{TompsonSampling_hariri2014context}, the authors have proposed to use a multi-armed bandit algorithm in order to quickly adapt the recommendations to the contextual changes in the case of interactive recommendations in an online setting. The main focus of that work is to apply bandit algorithms in interactive recommender systems and the change point detection is essentially an add-on feature independent of the learned model. 

In other works, Markov Models have been used to model user's sequential behavior  \cite{HHMM2015, rendleFPMC2010, he2016fusing, rendleFPMC2010}. In \cite{HHMM2015} a Hierarchical Hidden Markov Model is used to implicitly model the hidden states as context and generate recommendations using the last inferred context of user. Although, same approaches are used to model the dynamics of user preference, our model is based on single hidden layer compared to the a Hierarchical hidden structure. Besides simplicity, our work extends Viterbi algorithm for decoding task and generates recommendations using latent factors learned from Emission probabilities. The simplicity of our model compared to \cite{HHMM2015} makes it more efficient in training time. There are also other recent approaches \cite{he2016fusing, rendleFPMC2010} that are similar to our approach in terms of their overall goal of modeling user dynamics, but different in terms of their methodology.

\section{Problem Definition and Background}

In many recommendation domains users interact with a collection of available items by various actions such as viewing, clicking, or selecting items. These user-item interaction sequences form the underlying observations in our HMM-based change detection and recommendation framework.  

\subsection{Change Points}

Let $\mathcal{U} = \{u_1, u_2, ..., u_N\}$ be a set of users and $\mathcal{I}$ = $\{i_1, i_2,$ $...$ $i_M\}$ set of items. For each user $u$ the list of his/her interactions is denoted by $I_u$ = $\langle i^{(1)},$ $i^{(2)}$ $,..., $ $i^{(T)}\rangle$. Each element $i^{(t)}$ in this list represents the interaction of user $u$ with item $i$ at sequential index $t$ and all of the items are ordered based on the time of interaction. 

A change point $\Lambda$ in the list of user interactions is a sequential index that partitions this list into $I_1 = \langle i^{(1)},... i^{(\Lambda)} \rangle$ and $I_2 = \langle i^{(\Lambda+1)},...i^{(T)} \rangle$. There is no limit for the number of change points and the resulting segments, but for the sake of simplicity we only consider the situation when one change point exists. The challenge here is to accurately detect the change point $\Lambda$. More precisely, $\Lambda$ should partition the sequence of user interaction in a way that maximizes intra-partition similarity and minimizes inter-partition similarity of $I_1$ and $I_2$. 


\subsection{Hidden Markov Models}

Hidden Markov Models (HMM) belong to the category of probabilistic models specifically used for modeling sequential data. Markov Chains \texttt{MC} are the simplest in this category which are based on following independence assumption known as Markov property. Given a set of discrete time-based variables $Y \in \{ y_1$, $y_2$, ...$y_n\}$, and a sequence of these variables $\langle Y_1$, ..., $Y_{t-1}$, $Y_t \rangle$, in a first-order Markov models the probability of the $Y_t$ after seeing the sequence depends only on the last observation $P(Y_t = y|Y_1$, ..., $Y_{t-2}$, $Y_{t-1})$ = $P(Y_{t} = y|Y_{t-1})$. In recommendation domain due to the large number of items and hence the resulting data sparsity, Markov Chain models tend to perform poorly. Higher-order Markov Chains can be used, but at the cost of significantly higher time and space complexity.

Hidden Markov Models \texttt{HMM}s are the extensions of \texttt{MC} models that target the sequential pattern in the data using the transition probabilities between the hidden states instead of observations. The assumption behind hidden states $Z \in \{z_1$, ..., $z_k\}$ is that the transition probabilities cannot be inferred directly using observations. Formally, an HMM model is defined by an initial hidden state distribution $\pi$, state transition probabilities $\mathcal{A} = P(Z_t|Z_{t-1})$, and emission probabilities $\mathcal{B} = P(Y_t|Z_t)$. We, this, denote an HMM by $\theta = (\mathcal{A}, \mathcal{B}, \pi)$.   

In the literature, HMMs have been used to solve three different general problems that make these models very practical and suitable in many situations \cite{gales2008application}.
\begin{enumerate}
\item \textbf{Likelihood estimation problem}: Given a sequence of observations $\langle Y_1$, ..., $Y_{t-1}$, $Y_t \rangle$ , and $\theta$, determine the $P(Y_1$, ...$Y_t|\theta)$.
\item \textbf{Decoding problem}: Given a sequence of observations and $\theta$, what is the most likely sequence of hidden states.
\begin{equation*}
    \operatorname*{arg\,max}_{Z_1,...Z_t}{P(Z_1,...,Z_t,Y_1,...,Y_t|\theta)}
\end{equation*}
\item \textbf{Learning problem}: Given a sequence of observations and a set of hidden states, learn HMM parameters $\theta$ using maximum likelihood estimation.
\end{enumerate}

In the following section we demonstrate how we can adapt these problems and their solutions to the problem of detecting changes in user preferences over time. The identified change points can then be used to generate recommendations. 

\section{Proposed Framework}

\subsection{Change Point Detection}

Given a sequence of user-item interactions (implicitly representing user's preferences on items), we first learn $\theta = (\mathcal{A}, \mathcal{B}, \pi)$. We use the well-known \textit{Baum-Welch} algorithm to learn the model from the data \cite{BaumWelch1986}. This method is based on the Expectation-Maximization (EM) algorithm to find the maximum likelihood estimate of $\theta$ using the sequence of observations. Next, using the learned model we ``decode" the hidden states related to each observation (i.e., each item in the interaction sequence). The standard algorithm for this task is the \textit{Viterbi} algorithm \cite{viterbi}. Given a sequence of observations and $\theta$, \textit{Viterbi} algorithm uses a dynamic programming approach to find the the most likely hidden states corresponding to the sequence of observations. It is this generated sequence of hidden states which represents the dynamics of user preferences over time. 

In order to identify the change point $\Lambda$ we use the \textit{Viterbi} algorithm as follows. \textit{Viterbi} relies on $\theta$ to find the maximum path $\mathcal{V} = \langle v_1,...,v_T\rangle$ among the hidden states corresponding to each item in the sequence of observations. Suppose that the hidden state corresponding to the \textit{Viterbi}'s path at time $t$ is denoted by $\mathcal{V}_t$. A change is detected if there is a hidden state $\mathcal{V}_t$ with maximum tendency to change from $\mathcal{V}_{t-1}$, that is:

\begin{equation}
\Lambda' = \operatorname*{arg\,max}_{t}{P(\mathcal{V}_t|\mathcal{V}_{t-1}) P(i_t|\mathcal{V}_t)}
\end{equation}
Where $\mathcal{V}_t \neq \mathcal{V}_{t-1}$.
 
The change point detection algorithm is specified below.

\begin{algorithm}
\caption{Hidden Markov Change point Detection HMCD}
\label{alg_hmm_change_detection}
\KwIn{Set of sequences of User-Item interactions $X = \{ I_{u_1},...,I_{u_n} \}$, Number of Change Points $k$, Number of Hidden States $h$.}
\KwOut{Partitioned User-Item Matrix $M$.}
    \vspace{0.8em}
    {$M$: User-item interaction matrix $M \in \{0, 1\}^{n'\times m}$, where $n' = n \times k$.}
    
    \tcc{Learn HMM Model.}
    {$\theta \gets \textit{Baum-Welch}(X, \hspace{0.2em} h)$}
    
    \For {$u \in \mathcal{U}$} {
    \tcc{Using \textit{Viterbi}, Decode each item $i \in I_u$.}
    {$\mathcal{V} \gets$ \hspace{0.2em} $Viterbi(\theta, \hspace{0.2em} I_u)$}
    
    \vspace{0.5em}
    
    \tcc{Find time index of top-$k$ maximum State Changes in $\mathcal{V}$'s path.}
    {$\Lambda' = \{ t | \mathcal{V}_t \neq \mathcal{V}_{t-1}\land \operatorname*{arg\,max}_{t}^{k}{P(\mathcal{V}_t|\mathcal{V}_{t-1}) P(i^t|\mathcal{V}_t)} \}$}
    
    \vspace{0.5em}
    \tcc{Partition $I_u$ into $k+1$ parts using $\Lambda'$.}
    {$\langle I^0_u,...,I^k_u \rangle = Partition(I_u, \hspace{0.2em} \Lambda')$}
    
    \For {$i=0$ \KwTo $k$}{
        $M_{u\times k+i} = I^i_u$
    }
    }
{return $M$}
\end{algorithm}

\subsection{Generating Recommendations Using Detected Change Points} \label{recommendation_approach}

In this section we present two approaches for generating recommendations integrating the identified change points in user preferences. The high-level flow diagram for the overall framework is depicted in Figure~\ref{fig:flow_diagram}.

\textbf{Sequence-based Matrix Factorization Recommendation SMF}: First method is using Matrix Factorization models based on change points. After decoding the hidden states of each item in $I_u$ and segmenting this sequence using Algorithm~ \ref{alg_hmm_change_detection}, we treat each segment of items as a user profile vector used in the factorization model. More formally, given $I_u$ the sequence of interactions of user $u$, the HMM decoding is used to identify a change point $\Lambda'$. Then assuming that there is only one change point, $I_u$ will be segmented into $I^1_u = \langle i^{(1)},... i^{(\Lambda')} \rangle$ and $I^2_u = \langle i^{(\Lambda'+1)},...i^{(T)} \rangle$. The new segments for every user $u$ are used to generate a new user-item interaction matrix $M$. To find approximate factorization of $M \approx p.q^T$, we use Non-negative matrix factorization. The objective in this method is to minimize the Euclidean distance between the approximated matrix $p.q^T$ and the actual matrix $M$. More precisely, NMF minimizes $|| M - p.q^T ||^2$ with respect to $p$ and $q$, subject to $p,q \geqslant 0$. Since, change point detection has decoupled the items of each user profile in $M$ which sequentially do not belong to the same hidden state, the association of items with sequential likelihood in matrix factorization will not be lost. Therefore, using NMF we can estimate $q^T$, the item factors of $M$, which are based on the sequential patterns of items in the original user-item matrix $X$. Note that as long as we keep the number of change points small, the sparsity as a result of segmentation should not decrease accuracy of recommendations compared to static matrix factorization. 

In order to generate recommendations for $u$, we use the last segment $k$ of user interactions $I^k_u$. We select the top 10 most similar items to all of the items in $I^k_u$ for recommendation. In other words, we score items for recommendation as follows:

\begin{equation}
    score(u, i) = 1/|I^k_u| \sum_{i' \in I^k_u}{ \sum_{i \in \Psi_{i'}}{\mathbbm{1}}}
\end{equation}

Where a simple matching is used to find top-$l$ most similar items to each item: $\Psi_j = \operatorname*{arg\,max}_{j'}^{l}{(q^T_j.q_{j'})}$.
\vspace{0.6em}

\textbf{HMM-Based Recommender (HMMR)}: The second recommendation method is solely based on hidden states which were learned by HMM models in the previous step. The goal is to find a sequential item factors, in order to recommend items based on the last state of user according to change points. In the HMM model using emission probabilities $\mathcal{B}$ , for each hidden state $s$ and item $i$ we have the probability $P(i|s)$. However, for each item we need a probability distribution over the feature space of hidden states as latent factors $P(s|i)$. Using Bayes rule, the estimation of $P(s|i)$ is as follows:

\begin{equation}
    P(s|i) = \frac{P(i|s) \times P(s)}{P(i)}
\end{equation}    
Where the normalized marginal distributions of item $i$ and state $s$ are as follows. 
\begin{equation*}
P(i) = \frac{\sum_{s}{P(i|s)P(s)}}{\sum_{i'}{\sum_{s'}{P(i'|s')P(s')}}} \text{\space{  } ,}
\end{equation*}

\begin{equation*}
P(s) = \frac{\sum_{s'}{P(s|s')}}{\sum_{s'}{\sum_{s''}{P(s''|s')}}} \text{\space{ }}
\end{equation*}

For each item $i$ its latent factors $P(s|i)$ can be used the with any similarity matching to the latest preference of user segment $I_{u}^{last}$ to rank the items and make recommendations accordingly. We apply the same approach for generating the recommendations used in the factorization model. 

\section{Experiments}

In order to evaluate the proposed methods, we have designed two separate experiments: one to evaluate the effectiveness of the HMM-based change point detection when compared to baseline CPD methods, and anther for measuring the impact of change point detection on recommendation effectiveness. 



We use two HMM models with various number of hidden states labeled as $S2$ and $S10$ which indicates HMM models with $S=2$ and $S=10$ hidden states. We run our experiments on both of the proposed models (explained in section 4) which are HMM-Based Recommender (HMMR) and Sequence-based Matrix Factorization SMF. Therefore, SMF-S10 represents the sequence-based matrix factorization with $S=10$ hidden states, and HMMR-S10 represents the HMM-based recommender with $S=10$ hidden states.

\subsection{Dataset}

We used the Spotify Playlist dataset for our experiments \footnote{\href{https://recsys-challenge.spotify.com/}{RecSys Challenge 2018.}}. We randomly sampled 6,000 playlists from 1 million playlists available in this dataset which have been created by Spotify's users. Our sample contained 6,905 unique Albums, 475,838 playlist-album pairs with 98.9\% sparsity. Also, the average length of playlists was $\approx80$. We chose this dataset because average length playlists are usually focused on one or two moods or genres. Therefore, this characteristic of playlists make them a good candidate for concatenating multiple playlists as a way to simulate change points in the data. The procedure for generating the data is as follows. First, we randomly sampled two playlists $p_1$ and $p_2$ from the pool of 1M playlists. Then, we combined them by selecting a random size window of each playlist. The ground truth for change point is specified by the point where $p_1$ and $p_2$ are concatenated. We use all of the users to measure the accuracy of change point detection task. For the recommendation, we hold the last 10 items of the playlist $p_2$ (in the same order) as testing data for evaluation.    



\subsection{Change Detection Baselines}

We compared the HMM-based change point detection algorithm to several standard change point detection methods often used in time series analysis.

\textbf{Cumulative Sum, CUSUM}: Cumulative Sum is one of the best known approaches for change detection in time series. For each item $i \in I_u$ = $\langle i^{(1)},$ $i^{(2)}$ $,..., $ $i^{(T)}\rangle$, CUSUM computes the $S_j = \sum_{t=0}^{j}{i^{(t)}}$. Whenever the cumulative sum exceeds a threshold value $S_j > \tau; \forall  j \leqslant T$, this method identifies $\Lambda' = j$ as the change point in the $I_u$. In order to tune the threshold parameter $\tau$, we started with average value of CUSUM among all item sequences and empirically tune  this parameter where the minimum $\Delta$ has been reached.

\textbf{Sliding Window, SW}:
The sliding partition point (change point) slides from the beginning of user interaction list $I_u$ until it reaches to a point $\Lambda'$ which maximizes intra-partition similarity and minimizes inter-partition similarity. Although, this is a greedy approach for the mentioned objective, in practice it would produce fairly accurate results. We use Euclidean distance as a measure of dissimilarity between every pair of items. 


\textbf{Random Partition, RP}:
This is a basic baseline that randomly assign a value to $\Lambda'$ such that $0 \leqslant \Lambda' \leqslant |I^l_u|$.  


In all these cases and for the HMM-based method, we paired the change point detection approach with the matrix factorization model described in the previous section. The goal was to determine which change point detection method works best and also to evaluate the impact of change point detection on effectiveness of recommendations in general.

To summarize, the change point detection baselines are as follows:

\begin{itemize}
    \item Cumulative Sum with NMF (CUSUM-NMF)
    \item Sliding Window with NMF (SW-NMF)
    \item Random Partition with NMF (RP-NMF)
\end{itemize}

All of the above methods use the same recommendation approach that has been discussed in subsection ~\ref{recommendation_approach}.

\subsection{Recommendation Baselines}

In our experiments, we used the following baseline recommendation methods to evaluate the effectiveness of proposed HMM-based recommendation framework when compared to traditional non-sequential approaches. In all of the matrix factorization approaches we set the number latent factor to 40 in our experiments. Empirically, this value produced the best performing model on our data.

\noindent\textbf{BPR-MF} \cite{bprrendle2009}: Bayesian Personalized Ranking optimizes a pairwise ranking objective function via stochastic gradient descent. 

\noindent\textbf{NMF} \cite{NMFlee2001algorithms}: Non-negative Matrix Factorization is another baseline method that decomposes the multivariate data for generating recommendations. NMF uses multiplicative algorithm in order to minimize the least squared error of predictions.

\noindent\textbf{PopRank}: This method generates a recommendation list for all users by ranking the items based on their popularity among users. Popularity of items in the data are defined by their frequency of being seen/rated by users.

\begin{table}[tbp]
  \centering
  \begin{tabular}{lc}
    \toprule
    Method& $\Delta = |\Lambda - \Lambda'|$ \\
    \midrule
     Sliding Window & 24.185 \\
     CUSUM & 21.871 \\
     Random Partition & 26.142 \\
     HMCD-S10 \space{       } & \textbf{17.147} \\
     HMCD-S2 & \textbf{16.314} \\
    \bottomrule
  \end{tabular}
  \caption{Error of Change Point Detection Methods}
  \label{tab:CH-Acc}
\end{table}

\subsection{Evaluation Metrics}

We used two types of metrics for the evaluation of our methods. First metric is for measuring the accuracy of change point detection. The error of a single change point prediction is defined by $\Delta = |\Lambda - \Lambda'|$ where $\Lambda$ is the ground truth change point in user sequence of interactions and $\Lambda'$ is the prediction.  As noted before, in our experiments we concatenated different playlists to simulate a change point. Thus the end of one playlist and the beginning of another represent the ground truth change points in our test data.

Second types of metrics are for measuring the ranking accuracy of recommendations. We use Precision, Recall, and normalized discounted cumulative gain (NDCG) for this purpose. The discounted cumulative gain is time-aware which means the gain for items selected earlier by a user are larger when compared to later items. In other words, the true ranking of items are based on the time-based non-decreasing order of items in a user profile.

\begin{figure}
    \centering
    \includegraphics[scale=0.34]{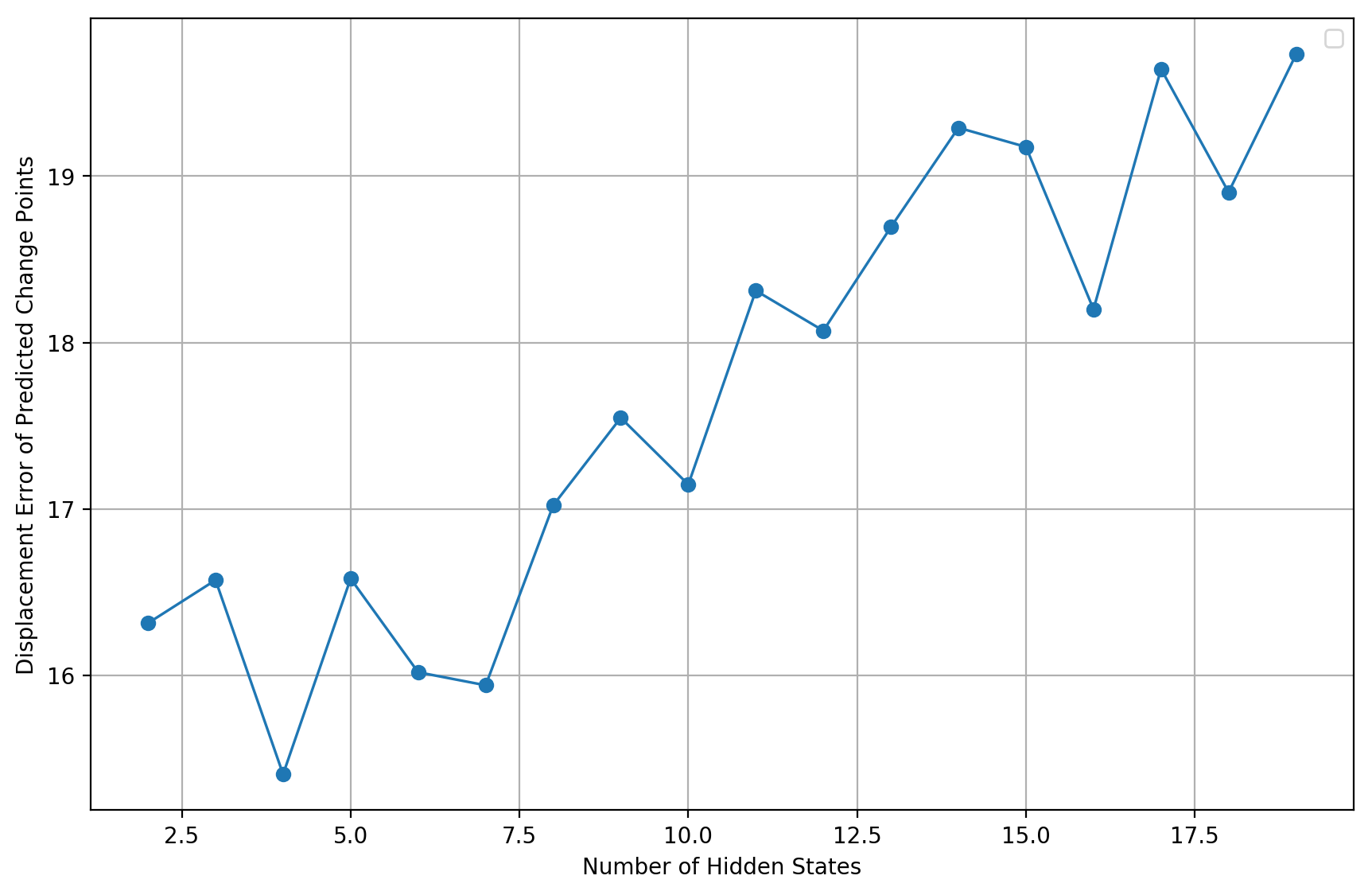}
    \caption{Error of predicted change points using different number of hidden states in HMM.}
    \label{fig: HMM_state_ch_acc}
\end{figure}

\begin{figure}
    \centering
    \includegraphics[scale=0.35]{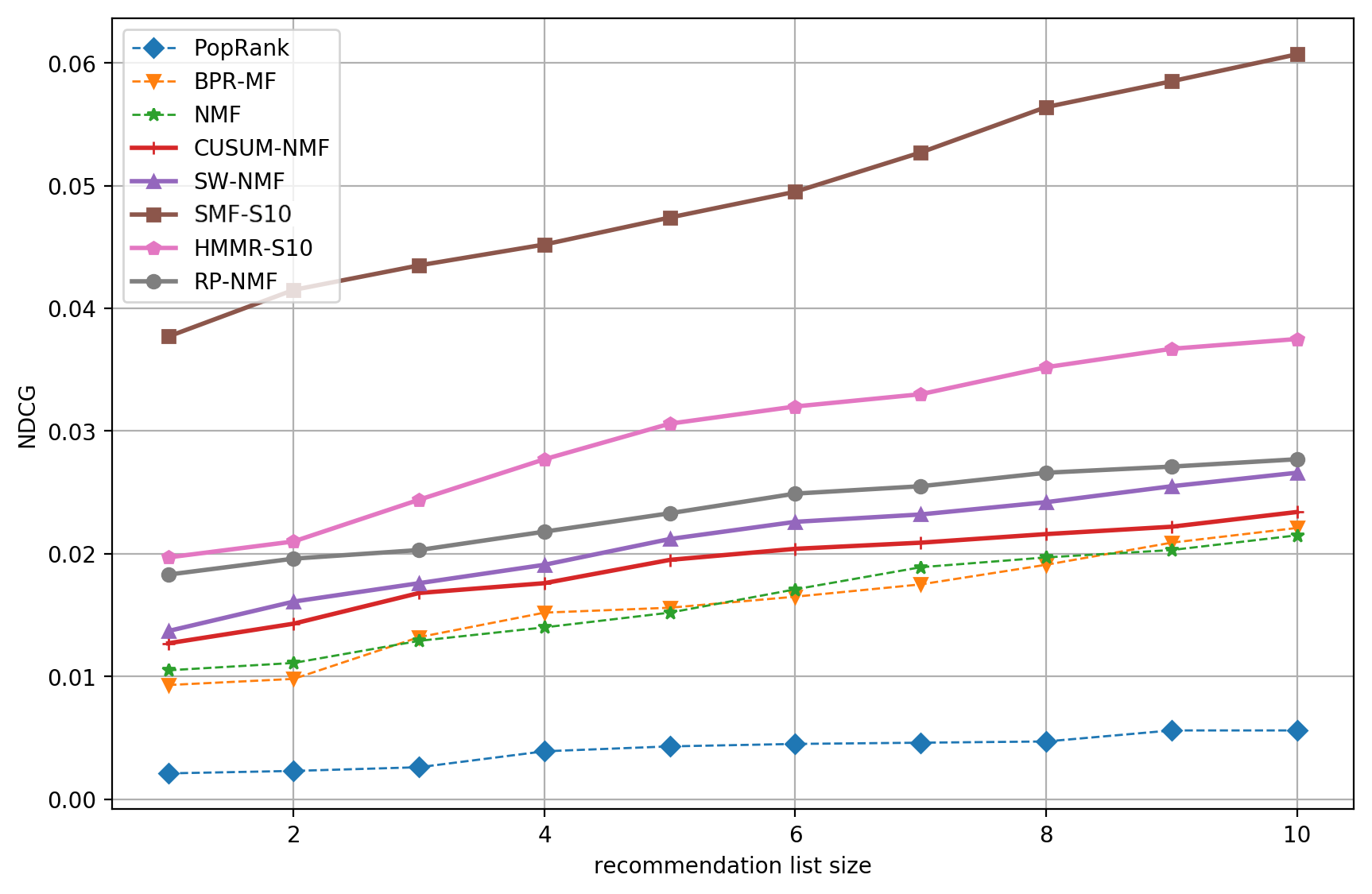}
    \caption{Comparison of Models based on NDCG}
    \label{fig: all_models_ndcg}
\end{figure}

\begin{figure}
    \centering
    \includegraphics[scale=0.35]{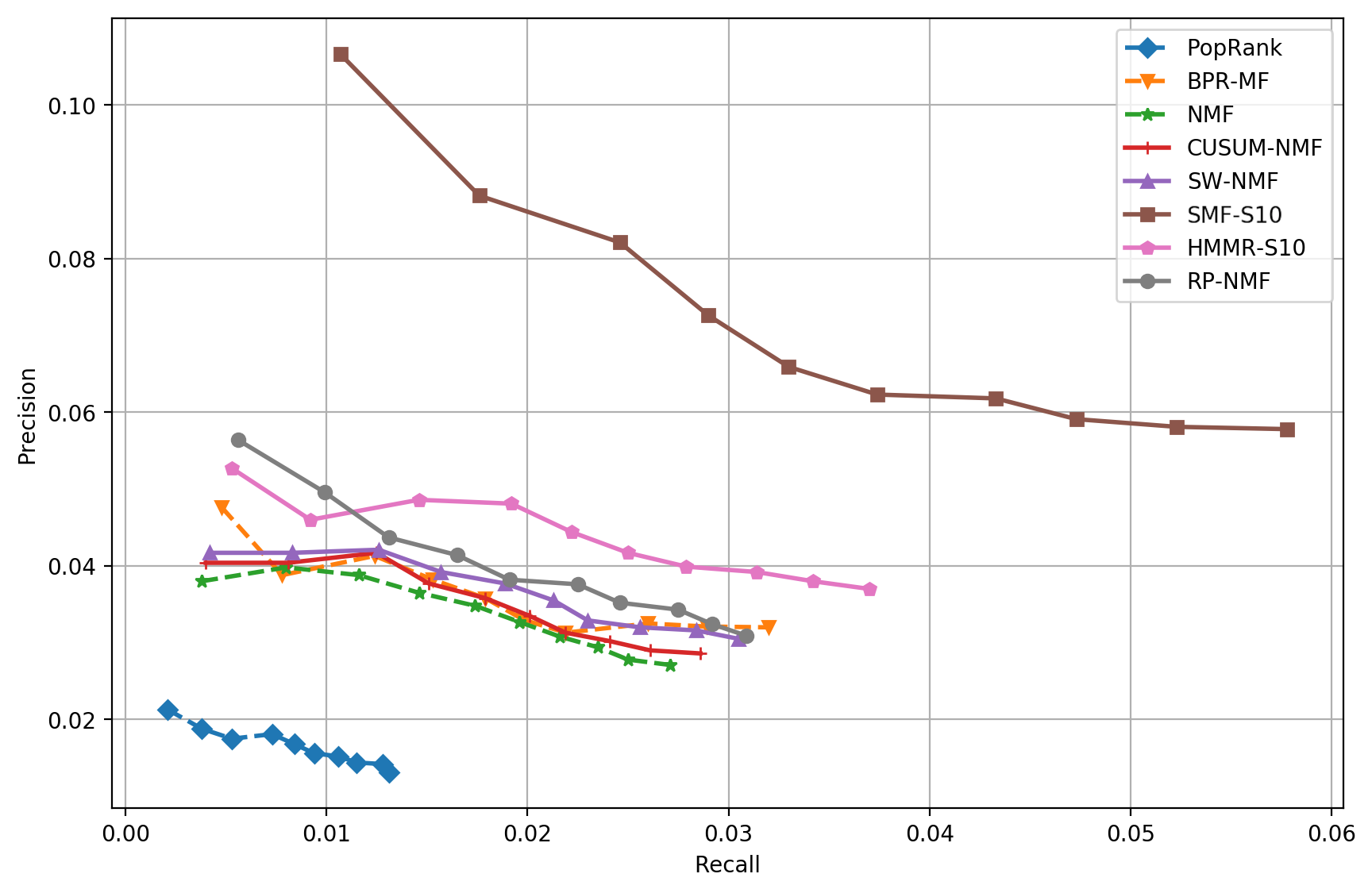}
    \caption{Comparison of Models based on Precision-Recall curves.}
    \label{fig: all_models_prec-recall}
\end{figure}

\subsection{Experimental Results}

The first task in our experiments was to evaluate the accuracy of change point detection. Table~\ref{tab:CH-Acc} shows the displacement error of change point detection using the baseline methods versus HMCD. We trained two HMM models with 2 and 10 hidden states and labeled them as HMCD-S2 and HMCD-S10, respectively. Since on average the length of each mixed playlist is about 80, the chances of a random prediction hitting the true change point $\Lambda$ would be $1/80$. The displacement error $\Delta$ of this approach should be larger. However, when we were generating the change points using playlists, we sampled $\Lambda$ from a uniform distribution. Therefore, the results of RP were better than a completely random guess. 

HMCD-S2 (HMM with 2 hidden states) and HMCD-S10 both outperform the baseline methods. An interesting experiment after this observation would be to examine the impact of increasing the number of HMM hidden states on $\Delta$. We illustrate this result in Figure~\ref{fig: HMM_state_ch_acc}. As can be seen in this figure, increasing the number of hidden states increases the error $\Delta$ in this dataset. The reason might be the number of change points we devised in the process of generating mixed playlists and also the the uniqueness of genre or topic of combined playlists.

Figure~\ref{fig: all_models_ndcg} and figure~\ref{fig: all_models_prec-recall} show the results of evaluating the accuracy of recommendations in terms of NDCG and Precision-Recall curves. In these figures, the static methods that do not model the dynamics of user preferences are illustrated by dashed lines. In our results, these models clearly do not perform as well as the sequential models. Also, it can be seen in both experiments that SMF with 10 hidden states outperforms all the other methods. The HMMR method is the second best model in our experiments. These results emphasize the importance of change detection and effectiveness of our methods for change detection and recommendation.

\section{Conclusion}

In this work, we introduced the concept of change point detection in order to model the dynamics of user preferences in sequential recommendation tasks. We devised a recommendation framework based on Hidden Markov Models to detect changes in user preferences and use the identified change points to appropriately target generated recommendations. Specifically we extended the standard Viterbi algorithm to detect the change points in sequences of user-item interactions.

The recommendation framework can use the identified change points in two ways. In one approach the change points can be used in a pre-filtering step to segment user profiles and feed the segmented data into a standard non-sequential recommendation model such as matrix factorization. In the second, more integrated approach, the emission probabilities from the HMM are used to directly generate recommendations based on the representation of items as a probability distribution over hidden states of HMM. Using Spotify playlist dataset, we demonstrated the effectiveness of our methods in terms of both accuracy of change point detection and accuracy of recommendations.

In future work, we plan to extend our experiments with other datasets involving different types of user-item interactions. Also, we will examine the use of change point detection with other recommendation models. Furthermore, we will explore the effectiveness of change point detection in situations where both short-term and long-term preferences need to be considered in recommendation generation.

\bibliographystyle{ACM-Reference-Format}
\bibliography{paper}

\end{document}